\documentclass{article}
\usepackage{amssymb,amsfonts,amsmath,stmaryrd}
\usepackage{indentfirst}
\usepackage{enumerate,float}
\usepackage{color}
\usepackage{tikz}
\usetikzlibrary{arrows,snakes,backgrounds}
\usepackage{ytableau}
\usepackage{physics}
\usepackage{comment}
\usepackage[hidelinks,colorlinks=true,unicode]{hyperref}
\hypersetup{linkcolor=orange, citecolor=orange, filecolor=orange, urlcolor=orange}
\usepackage{url}
\usepackage[natbib=true, style=numeric-comp, sorting=none, maxbibnames=10]{biblatex}
\def\be{\begin{eqnarray}}
\def\ee{\end{eqnarray}}
\def\nn{\nonumber}

\def\p{\partial}

\def\yt{\scriptsize \text }

\def\H{\omega}

\definecolor{red}{rgb}{1,0,0}
\definecolor{orange}{rgb}{1,0.5,0}
\definecolor{violet}{rgb}{0.7,0,1}

\hoffset= - 1.0in         

\begin{document}
\title{{\Large {\bf Hunt for 3-Schur polynomials}
\vspace{.1cm}}
\author{
{\bf A.~Morozov $^{a,b,c,d}$},
{\bf N.~Tselousov $^{a,b,d}$} \date{ }}
}
\maketitle

\vspace{-5cm}

\begin{center}
\hfill MIPT/TH-13/22

\hfill ITEP/TH-16/22

\hfill IITP/TH-14/22
\end{center}

\vspace{2.5cm}

\begin{center}
\begin{small}
$^a$ {\it MIPT, 141701, Dolgoprudny, Russia}\\
$^b$ {\it NRC “Kurchatov Institute”, 123182, Moscow, Russia}\\
$^c$ {\it IITP RAS, 127051, Moscow, Russia}\\
$^d$ {\it ITEP, Moscow, Russia}\\
\end{small}
\end{center}

\vspace{0.5cm}

 
\begin{abstract}
{\footnotesize
This paper describes our attempt to understand the recent success
of Na Wang in constructing the 3-Schur polynomials, associated with the plane partitions.
We provide a rather detailed review and try to figure out the new insights,
which allowed to overcome  the problems of the previous efforts.
In result we provide a very simple definition of time-variables ${\bf P}_{i\geqslant j}$
and the cut-and-join operator $\hat W_2$, which generates the set of $3$-Schur functions.
Some coefficients in $\hat W_2$ remain undefined and require more effort to be fixed.
}
\end{abstract}

\section{Introduction}

Schur functions play a prominent role in modern theoretical physics.
Today we understand the basic reason:
they provide the right non-trivial basis, adjusted to exhaustive description
of Gaussian integrals \cite{Mironov:2022fsr,Mironov:2017och,Mishnyakov:2022bkg, Wang:2022fxr,Bawane:2022cpd, Cassia:2020uxy}.
Somewhat unexpectedly,
this basis includes an {\it a priori} hidden structure of Young diagrams
(partitions) and a good piece of representation theory behind them.
Schur functions have a number of generalizations --
to Jack, Q-Schur \cite{Mironov:2021wfh}, Macdonald \cite{Macdonald,Mironov:2019uoy} and Shiraishi \cite{Awata:2020xfq} functions,
as well as a more mysterious one to Kerov functions \cite{Kerov, Ker1Mironov:2019, Ker2Mironov:2018nie}.
Generalization to Jacks is the most straightforward and associated to
$\beta$-deformed Gaussians.
Macdonald functions, while formally associated with the further $q,t$-deformation,
are long anticipated to be also related to plane partitions,
i.e. to the still undeveloped theory of 3-Schur functions
\cite{Morozov:2018fjb}.
The lasting failure to develop their theory by elementary means,
suggests the full-fledged use of the underlying representation theory,
which in this case is the theory of Yangians \cite{Maulik:2012wi, Shiffmann, Galakhov:2020vyb, Galakhov:2021xum}
(and further raised to DIM algebras \cite{Mironov:2016yue, Ghoneim:2020sqi, Awata:2018svb}).
Recently, Na Wang  \cite{Wang1} made a very good progress in this direction,
based on the old presentations  of \cite{Tsymbaliuk:2014fvq, Prochazka:2015deb} and further developed in
\cite{Wang2,Wang3,Wang4}.
Our goal in this paper is to reproduce and understand these results --
and to figure out new insights, which allowed to overcome the problems,
reported in \cite{Morozov:2018fjb, Morozov:2018fga, Morozov:2018lsn, Zenkevich:2017tnb}.

\section{The case of $2$-Schur and Jack polynomials}

To outline the suggested strategy, we present the sample calculation
for Jack polynomials.
The ordinary $2$-Schur functions arise as a particular case when parameter $\beta=1$. We do not describe the origins of the method, which is inspired by the above-cited literature.
It will be discussed in greater details in a lengthier presentation in future works.

The starting point is the {\it cut-and-join} operator 
\be
\label{cut and join Jack}
\hat W_2 = \frac{1}{2}\sum_{a,b=1}^{\infty} \left( a \, b \,{\bf p_{a+b}} \, \frac{\p^2}{\p {\bf p_a}\p {\bf p_b}}
+ (a+b) \, {\bf p_a p_b}\frac{\p}{\p {\bf p_{a+b}}}\right)
- \frac{\eta^2-1}{2\eta}\sum_{a=2}^{\infty} a(a-1){\bf p_a}\frac{\p}{\p {\bf p_a}}
\ee
where we rescaled \footnote{Comparing with ordinary presentation (4.83) from \cite{Prochazka:2015deb}. Note the change of variable $h=\eta$.} the time-variables ${\bf p_a} := \eta p_a$ with $\boxed{\eta:=\sqrt{\beta}}$
to make the formulas more symmetric and transparent.
In particular, the $\beta$-dependent parameter is the one which defines the cental charge
in the AGT relations $c=1-\frac{6(\eta^2-1)^2}{\eta^2}$.
This will also provide an unusual, but clever normalization of Jacks.

In case of Jack polynomials the cut-and-join operator \eqref{cut and join Jack} is known in full generality therefore one can derive all the Jack polynomials as its eigenfunctions. However, we develop another method of deriving Jacks from cut-and-join operator that will be useful in lifting the whole constuction to the level of 3-Schur functions.

Our method involves operators $\hat e_0$ and $\hat e_1$ that are defined as follows:
\begin{align}
    \hat e_0 &:= {\bf p_1} \\
    \hat e_1 &:= \left[ \hat W_2,\hat e_0 \right]= \sum_{a=1}^{\infty} a \, {\bf p_{a+1}}\frac{\p}{\p {\bf p_a}}
\end{align}
The action of these operators 
on the vacuum state $\ket{\varnothing} := 1$   is independent of $\eta$,
\begin{align}
\begin{aligned}
\hat e_0 \cdot 1 &= {\bf p_1}, &\hspace{5mm} \hat e_1 \cdot 1 &= 0 \\
\hat e_0 \hat e_0 \cdot 1 &= {\bf p_1^2}, &\hspace{5mm} \hat e_1\hat e_0\cdot 1 &={\bf p_2} \\
\hat e_0 \hat e_0 \hat e_0 \cdot 1 &= {\bf p_1^3}, &\hspace{5mm} \hat e_1\hat e_0 \hat e_0\cdot 1 &= 2{\bf p_2p_1}, &\hspace{5mm}
\hat e_0\hat e_1\hat e_0\cdot 1 &=  {\bf p_2p_1}
\end{aligned}
\end{align}
This $\eta$-independence depends on the  factor $(a-1)$ in the last term in $\hat W_2$,
which could actually be $a-c$ with arbitrary $c$ -- what corresponds to linear
combination of $\hat W_2$ with $\hat W_1:=\sum_{a=1} a {\bf p_a}\frac{\p}{\p {\bf p_a}}$.
We prefer our choice with $\eta$-independent $\hat e_1$.

Our method also involves the content-function $\omega_{i,j}$ for the Young diagrams:
\begin{equation}
    \boxed{\omega_{i,j} := -\eta(i-1) + \frac{(j-1)}{\eta} }
\end{equation}
We use the convention that the starting box of the Young diagram has coordinate $(1,1)$, therefore $\omega_{1,1} = 0$.\\
We provide illustrating examples of computation of the first three levels of Jacks polynomials and then explain the construction in great detail.

$\bullet$ 1 level:
\be
\ytableausetup{boxsize = 0.5em}
\hat e_0 \cdot 1 = {\bf J}_{\begin{ytableau} \ \end{ytableau}} = {\bf p_1} 
\ee

$\bullet$ 2 level:
\begin{align}
\ytableausetup{boxsize = 0.5em}
    \begin{pmatrix}
    \hat e_0 \hat e_0 \cdot 1 \\
    \\
    \textcolor{red}{\hat e_1} \hat e_0 \cdot 1 \\
    \end{pmatrix} = 
    \begin{pmatrix}
    \underbrace{C_{\begin{ytableau} \yt{1} & \yt{2}  \\ \end{ytableau}}}_{1} & 
    \underbrace{C_{\begin{ytableau} \yt{1} \\ \yt{2} \\ \end{ytableau}}}_{1} \\
    \textcolor{red}{\omega_{1,2}} C_{\begin{ytableau} \yt{1} & \textcolor{red}{\yt{2}} \\ \end{ytableau}} & 
    \textcolor{red}{\omega_{2,1}} C_{\begin{ytableau} \yt{1} \\ \textcolor{red}{\yt{2}} \\ \end{ytableau}} \\
    \end{pmatrix}
    \begin{pmatrix}
    {\bf J}_{\begin{ytableau} \ & \ \end{ytableau}} \\
    \\
    {\bf J}_{\begin{ytableau} \ \\ \ \end{ytableau}} \\
    \end{pmatrix}
\end{align}

\be
{\bf J}_{\begin{ytableau} \ & \ \end{ytableau}} = \frac{\eta}{\eta^2+1}(\eta{\bf p_1^2} + {\bf p_2}) \nn \\
{\bf J}_{\begin{ytableau} \ \\ \ \end{ytableau}} = \frac{1}{\eta^2+1}({\bf p_1^2} - \eta{\bf p_2}) 
\ee

$\bullet$ 3 level:
\begin{align}
\ytableausetup{boxsize = 0.5em}
    \begin{pmatrix}
    \hat e_0 \hat e_0 \hat e_0 \cdot 1 \\
    \\
    \hat e_0 \textcolor{red}{\hat e_1} \hat e_0 \cdot 1 \\
    \\
    \textcolor{red}{\hat e_1} \hat e_0 \hat e_0 \cdot 1 \\
    \\
    \textcolor{red}{\hat e_1} \textcolor{red}{\hat e_1} \hat e_0 \cdot 1 \\
    \end{pmatrix} = 
    \begin{pmatrix}
    \underbrace{C_{\begin{ytableau} \yt{1} & \yt{2} & \yt{3} \\ \end{ytableau}}}_{1} & 
    \underbrace{C_{\begin{ytableau} \yt{1} & \yt{2} \\ \yt{3} \\ \end{ytableau}} + 
    C_{\begin{ytableau} \yt{1} & \yt{3}  \\ \yt{2} \end{ytableau}}}_{2} & 
    \underbrace{C_{\begin{ytableau} \yt{1} \\ \yt{2}  \\ \yt{3} \end{ytableau}}}_{1} \\
    \textcolor{red}{\omega_{1,2}} C_{\begin{ytableau} \yt{1} & \textcolor{red}{\yt{2}} & \yt{3} \\ \end{ytableau}} & 
    \textcolor{red}{\omega_{1,2}} C_{\begin{ytableau} \yt{1} & \textcolor{red}{\yt{2}} \\ \yt{3} \\ \end{ytableau}} + 
    \textcolor{red}{\omega_{2,1}} C_{\begin{ytableau} \yt{1} & \yt{3}  \\ \textcolor{red}{\yt{2}} \end{ytableau}} & 
    \textcolor{red}{\omega_{2,1}} C_{\begin{ytableau} \yt{1} \\ \textcolor{red}{\yt{2}}  \\ \yt{3} \end{ytableau}} \\
    \textcolor{red}{\omega_{1,3}} C_{\begin{ytableau} \yt{1} & \yt{2} & \textcolor{red}{\yt{3}} \\ \end{ytableau}} & 
    \textcolor{red}{\omega_{2,1}} C_{\begin{ytableau} \yt{1} & \yt{2} \\ \textcolor{red}{\yt{3}} \\ \end{ytableau}} + 
    \textcolor{red}{\omega_{1,2}} C_{\begin{ytableau} \yt{1} & \textcolor{red}{\yt{3}}  \\ \yt{2} \end{ytableau}} & 
    \textcolor{red}{\omega_{3,1}} C_{\begin{ytableau} \yt{1} \\ \yt{2}  \\ \textcolor{red}{\yt{3}} \end{ytableau}} \\
    \textcolor{red}{\omega_{1,2}\omega_{1,3}} C_{\begin{ytableau} \yt{1} & \textcolor{red}{\yt{2}} & \textcolor{red}{\yt{3}} \\ \end{ytableau}} & 
    \textcolor{red}{\omega_{1,2} \omega_{2,1}} \left( C_{\begin{ytableau} \yt{1} & \textcolor{red}{\yt{2}} \\ \textcolor{red}{\yt{3}} \\ \end{ytableau}} + 
     C_{\begin{ytableau} \yt{1} & \textcolor{red}{\yt{3}}  \\ \textcolor{red}{\yt{2}} \end{ytableau}} \right) & 
    \textcolor{red}{\omega_{2,1} \omega_{3,1}} C_{\begin{ytableau} \yt{1} \\ \textcolor{red}{\yt{2}}  \\ \textcolor{red}{\yt{3}} \end{ytableau}} \\
    \end{pmatrix}
    \begin{pmatrix}
    {\bf J}_{\begin{ytableau} \ & \ & \ \end{ytableau}} \\
    \\
    {\bf J}_{\begin{ytableau} \ & \ \\ \ \end{ytableau}} \\
    \\
    {\bf J}_{\begin{ytableau} \ \\ \ \\ \ \end{ytableau}} \\
    \end{pmatrix}
    \label{Jacksfromstates}
\end{align}

\begin{align}
\ytableausetup{boxsize = 0.5em}
{\bf J}_{\begin{ytableau} \ & \ & \ \end{ytableau}} &= \frac{\eta^2}{(\eta^2+2)(\eta^2+1)}(\eta^2{\bf p_1^3} +3\eta{\bf p_2p_1}+ 2{\bf p_3})
\hspace{10mm} &C_{\begin{ytableau} \yt{1} & \yt{3} \\ \yt{2} \\ \end{ytableau}} = \frac{2(\eta^2+2)}{3(\eta^2+1)}, \nn \\
{\bf J}_{\begin{ytableau} \ & \ \\ \ \end{ytableau}} &=  \frac{3\eta}{(\eta^2+2)(2\eta^2+1)} (\eta{\bf p_1^3}  +(1-\eta^2)\bf p_2\bf p_1 -\eta{\bf p_3}) \hspace{10mm} &C_{\begin{ytableau} \yt{1} & \yt{2} \\ \yt{3} \\ \end{ytableau}} =\frac{2(2\eta^2+1)}{3(\eta^2+1)} \label{3 level Jacks} \\
{\bf J}_{\begin{ytableau} \ \\ \ \\ \ \end{ytableau}} &=   \frac{1}{(2\eta^2+1)(\eta^2+1)}({\bf p_1^3}-3\eta{\bf p_2p_1} + 2 \eta^2{\bf p_3}) \nn
\end{align}

\subsection{Explanation of the method}
\label{Explanation of the method}
In our method Jack polunomials are solutions of the linear system of equations. This system is constructed as follows:
\begin{enumerate}
    \item The l.h.s. of the equation is the column of "words" of operators $\hat e_0, \hat e_1$ acting on $\ket{\varnothing} := 1$. On the level $n$ each "word" consist of $n$ operators. Note that $\hat e_1 \cdot 1 = 0$.
    \item The r.h.s of the equation on the level $n$ is the matrix $M_n$ multiplied by the column of Jack polynomials.
    \item The matrix elements of $M_n$ consist of coefficients $C_{\bar R}$, where ${\bar R}$ is a Young tableaux, and combinations of content-function computed at specific points. Coefficients $C_{\bar R}$ enter the column of the matrix that corresponds to Jack polynomial ${\bf J}_{R}$ of Young diagram $R$.
    \item The rule to insert content-function is as follows. We look at the positions $k$ of $\textcolor{red}{\hat e_1}$ in the "word" -- counted from the right.
    Then we look what is the box $\textcolor{red}{\Box_k}$ in the diagram labeled by this number and put the corresponding factor $\textcolor{red}{\omega_{\Box_k}}$ in front of the factor  $C_{\bar R}$. The procedure depends on $k$, i.e. on the "word" at the l.h.s. and on the Young tableaux $\bar R$ at the r.h.s. For example, for $\hat e_0\textcolor{red}{\hat e_1} \hat e_0 \cdot 1$ we have a single $k=2$, thus the coefficient in front of   $C_{\begin{ytableau} \yt{1} & \textcolor{red}{\yt{2}} \\ \yt{3} \end{ytableau}}$ will be $\textcolor{red}{\omega_{1,2}}$, while that in front of $C_{\begin{ytableau} \yt{1} & \yt{3} \\ \textcolor{red}{\yt{2}} \end{ytableau}}$ will be $\textcolor{red}{\omega_{2,1}}$. For the word $\textcolor{red}{\hat e_1 \hat e_1} \hat e_0 \cdot 1$ there will be two $k=2,3$, and both $C_{\begin{ytableau} \yt{1} & \textcolor{red}{\yt{2}} \\ \textcolor{red}{\yt{3}} \end{ytableau}}$ and $C_{\begin{ytableau} \yt{1} & \textcolor{red}{\yt{3}} \\ \textcolor{red}{\yt{2}} \end{ytableau}}$ acquire the same coefficients $\textcolor{red}{\omega_{2,1}\omega_{1,2}}$.
    \item The "word" in the first line does not contain $\hat e_1$ operators, therefore according to p.4 the matrix element corresponding to ${\bf J}_R$ is simply the sum of coefficients $C_{\bar R}$ for all possible Young tableau of shape $R$. To fix the normalization of Jack polynomials we choose the following condition: the sum of coefficients $C_{\bar R}$ is equal to the number of Young tableau of shape $R$.
    \begin{equation}
        \sum_{\bar R} C_{\bar R} = \text{number of Young tableau of shape $R$}
    \end{equation}
    In particular, for one line and one column Young diagram the corresponding coefficient is equal to one, because for these Young diagrams there are only one possible Young tableaux. For example, $C_{\begin{ytableau} \yt{1} & \yt{2} & \yt{3} \end{ytableau}} = 1$.
    \item We solve this system with respect to Jack polynomials ${\bf J}_R$ and unknown coefficients $C_{\bar R}$ expanding both sides of equations in a basis of times.
\end{enumerate}

One can check that the above defined Jack polynomials are eigenfunctions of $\hat W_2$ with the eigenvalues which depend only on the
diagram, not tableaux:
\be
\hat W_2\, {\bf J}_R =   \lambda_R {\bf J}_R, \ \ \ \ \ \ \ \ \ \
\lambda_R =  \sum_{\Box \in R} \omega_\Box
\label{Jackev}
\ee
We do not prove this statement here, however we show on a particular example, that the eigenfunction property is in full argeement with our definition of Jacks. Consider the following expression:
\be
\hat W_2 \hat e_0^3 \stackrel{(\ref{Jacksfromstates})}{=} \hat W_2{\bf J}_{\begin{ytableau} \ \\ \ \\ \ \end{ytableau}} +  2\hat W_2{\bf J}_{\begin{ytableau} \ & \ \\ \ \end{ytableau}} +  \hat W_2{\bf J}_{\begin{ytableau} \ & \ & \ \end{ytableau}}
\ee
that should be equal to
\be
\hat W_2{\bf J}_{\begin{ytableau} \ \\ \ \\ \ \end{ytableau}} +  2\hat W_2{\bf J}_{\begin{ytableau} \ & \ \\ \ \end{ytableau}} +  \hat W_2{\bf J}_{\begin{ytableau} \ & \ & \ \end{ytableau}}
\ \stackrel{(\ref{Jackev})}{=} \
 (\omega_{3,1}+\omega_{2,1}){\bf J}_{\begin{ytableau} \ \\ \ \\ \ \end{ytableau}}
+ 2(\omega_{2,1}+\omega_{1,2}){\bf J}_{\begin{ytableau} \ & \ \\ \ \end{ytableau}}
+ (\omega_{1,3}+\omega_{1,2}){\bf J}_{\begin{ytableau} \ & \ & \ \end{ytableau}}
\ee
according to the eigenfunction property. 
Both sides of this equation are indeed equal:
the r.h.s.
\be
 (\omega_{3,1}+\omega_{2,1}){\bf J}_{\begin{ytableau} \ \\ \ \\ \ \end{ytableau}}
+ 2(\omega_{2,1}+\omega_{1,2}){\bf J}_{\begin{ytableau} \ & \ \\ \ \end{ytableau}}
+ (\omega_{1,3}+\omega_{1,2}){\bf J}_{\begin{ytableau} \ & \ & \ \end{ytableau}}
\ \stackrel{(\ref{Jacksfromstates})}{=} \
(\hat e_1\hat e_0+\hat e_0\hat e_1)\hat e_0\cdot 1
\ee
while the l.h.s.
\be
\hat W_2 \hat e_0^3 \ = \
(\hat e_1\hat e_0+\hat e_0\hat W_2\hat e_0) \hat e_0 \cdot 1 \ = \
(\hat e_1\hat e_0+\hat e_0\hat e_1)\hat e_0\cdot 1
\ee
where we used the commutator $ [\hat W_2, \hat e_0] = \hat e_1$.

\subsection{Jacks at the level 4}

At the next level we encounter $5$ Young diagrams and $1+3+2+3+1=10$ Young tableaux.
Five polynomials ${\bf J}_R $ and  $10-5=5$ constants $C_{\bar R}$ are defined from
the $2^3=8$ decompositions:

{\normalsize
\begin{align}
\begin{aligned}
\ytableausetup{boxsize = 0.5em}
\hat e_0 \hat e_0 \hat e_0 \hat e_0 \cdot 1 &=  \underbrace{C_{\begin{ytableau} \yt{1} & \yt{2} & \yt{3} & \yt{4} \\ \end{ytableau}}}_{1} {\bf J}_{\begin{ytableau} \ & \ & \ & \ \end{ytableau}} + 
\underbrace{\left( C_{\begin{ytableau} \yt{1} & \yt{2} & \yt{3} \\ \yt{4} \\ \end{ytableau}} + 
    C_{\begin{ytableau} \yt{1} & \yt{2} & \yt{4} \\ \yt{3} \\ \end{ytableau}} + 
    C_{\begin{ytableau} \yt{1} & \yt{3} & \yt{4} \\ \yt{2} \\ \end{ytableau}}
    \right)}_{3}{\bf J}_{\begin{ytableau} \ & \ & \ \\ \ \end{ytableau}} 
    +\underbrace{ \left( C_{\begin{ytableau} \yt{1} & \yt{2} \\ \yt{3} & \yt{4} \\ \end{ytableau}} +
    C_{\begin{ytableau} \yt{1} & \yt{3} \\ \yt{2} & \yt{4} \\ \end{ytableau}}\right)}_{2} {\bf J}_{\begin{ytableau} \ & \ \\ \ & \ \end{ytableau}} + \\ 
    &+\underbrace{\left( C_{\begin{ytableau} \yt{1} & \yt{2} \\ \yt{3} \\ \yt{4} \\ \end{ytableau}} + 
    C_{\begin{ytableau} \yt{1} & \yt{3} \\ \yt{2} \\ \yt{4} \\ \end{ytableau}} + 
    C_{\begin{ytableau} \yt{1} & \yt{4} \\ \yt{2} \\ \yt{3} \\ \end{ytableau}}\right)}_{3}{\bf J}_{\begin{ytableau} \ & \ \\ \ \\ \ \end{ytableau}} + \underbrace{C_{\begin{ytableau} \yt{1} \\ \yt{2}  \\ \yt{3} \\ \yt{4} \end{ytableau}}}_{1}{\bf J}_{\begin{ytableau} \ \\ \ \\ \ \\ \ \end{ytableau}} \\
    \textcolor{red}{\hat e_1} \hat e_0 \hat e_0 \hat e_0 \cdot 1 &= 
    \textcolor{red}{\omega_{1,4}}C_{\begin{ytableau} \yt{1} & \yt{2} & \yt{3} & \textcolor{red}{\yt{4}} \\ \end{ytableau}} \, {\bf J}_{\begin{ytableau} \ & \ & \ & \ \end{ytableau}} + \left\{
    \textcolor{red}{\omega_{2,1}}C_{\begin{ytableau} \yt{1} & \yt{2} & \yt{3} \\ \textcolor{red}{\yt{4}} \\ \end{ytableau}} + 
    \textcolor{red}{\omega_{1,3}}\left(C_{\begin{ytableau} \yt{1} & \yt{2} & \textcolor{red}{\yt{4}} \\ \yt{3} \\ \end{ytableau}} + 
    C_{\begin{ytableau} \yt{1} & \yt{3} & \textcolor{red}{\yt{4}} \\ \yt{2} \\ \end{ytableau}}\right) \right\} {\bf J}_{\begin{ytableau} \ & \ & \ \\ \ \end{ytableau}} + 
    \left\{ \textcolor{red}{\omega_{2,2}}\left(C_{\begin{ytableau} \yt{1} & \yt{2} \\ \yt{3} & \textcolor{red}{\yt{4}} \\ \end{ytableau}} +
    C_{\begin{ytableau} \yt{1} & \yt{3} \\ \yt{2} & \textcolor{red}{\yt{4}} \\ \end{ytableau}}\right) \right\} {\bf J}_{\begin{ytableau} \ & \ \\ \ & \ \end{ytableau}} + \\ 
    &+\left\{ \textcolor{red}{\omega_{3,1}}\left(C_{\begin{ytableau} \yt{1} & \yt{2} \\ \yt{3} \\ \textcolor{red}{\yt{4}} \\ \end{ytableau}} + 
    C_{\begin{ytableau} \yt{1} & \yt{3} \\ \yt{2} \\ \textcolor{red}{\yt{4}} \\ \end{ytableau}}\right) + 
    \textcolor{red}{\omega_{1,2}}C_{\begin{ytableau} \yt{1} & \textcolor{red}{\yt{4}} \\ \yt{2} \\ \yt{3} \\ \end{ytableau}} \right\} {\bf J}_{\begin{ytableau} \ & \ \\ \ \\ \ \end{ytableau}} +
    \textcolor{red}{\omega_{4,1}}C_{\begin{ytableau} \yt{1} \\ \yt{2}  \\ \yt{3} \\ \textcolor{red}{\yt{4}} \end{ytableau}} \, {\bf J}_{\begin{ytableau} \ \\ \ \\ \ \\ \ \end{ytableau}}\\
\end{aligned}
\end{align}
}

We do not write the remaining six equations, they can be easily obtained by the rule,
formulated in the section \ref{Explanation of the method}.

The result is:
\begin{align}
\begin{aligned}
C_{\begin{ytableau} \yt{1} & \yt{2} & \yt{3} \\ \yt{4} \\ \end{ytableau}}  &= \frac{3(\eta^2+1)}{2(\eta^2+2)}, &\hspace{5mm}
C_{\begin{ytableau} \yt{1} & \yt{2} & \yt{4} \\ \yt{3} \\ \end{ytableau}} &= \frac{(2\eta^2+1)(\eta^2+3)}{2(\eta^2+2)(\eta^2+1)}, &\hspace{5mm}
C_{\begin{ytableau} \yt{1} & \yt{3} & \yt{4} \\ \yt{2} \\ \end{ytableau}} &= \frac{\eta^2+3}{2(\eta^2+1)};  \\
C_{\begin{ytableau} \yt{1} & \yt{2} \\ \yt{3} & \yt{4} \\ \end{ytableau}} &= \frac{2(2\eta^2+1)}{3(\eta^2+1)},& \hspace{5mm}
C_{\begin{ytableau} \yt{1} & \yt{3} \\ \yt{2} & \yt{4} \\ \end{ytableau}} &= \frac{2(\eta^2+2)}{3(\eta^2+1)}; & \\
C_{\begin{ytableau} \yt{1} & \yt{4} \\ \yt{2} \\ \yt{3} \\ \end{ytableau}} &= \frac{3(\eta^2+1)}{2(2\eta^2+1)}, &\hspace{5mm} 
C_{\begin{ytableau} \yt{1} & \yt{3} \\ \yt{2} \\ \yt{4} \\ \end{ytableau}} &= \frac{(\eta^2+2)(3\eta^2+1)}{2(2\eta^2+1)(\eta^2+1)}, &\hspace{5mm} 
C_{\begin{ytableau} \yt{1} & \yt{2} \\ \yt{3} \\ \yt{4} \\ \end{ytableau}} &= \frac{3\eta^2+1}{2(\eta^2+1)}.
\label{C constant level 4}
\end{aligned}
\end{align}

\begin{align}
\begin{aligned}
{\bf J}_{\begin{ytableau} \ & \ & \ & \ \end{ytableau}} &= \frac{\eta^3}{(\eta^2+3)(\eta^2+2)(\eta^2+1)} \left(\eta^3 {\bf p_1}^4+6\eta^2{\bf p_2p_1}^2+8\eta {\bf p_3p_1} + 3\eta {\bf p_2}^2 + 6{\bf p_4}\right) \\
{\bf J}_{\begin{ytableau} \ & \ & \ \\ \ \end{ytableau}} &= \frac{2\eta^2}{(\eta^2+3)(\eta^2+1)^2} \left( \eta^2{\bf p_1}^4+(3\eta-\eta^3) {\bf p_2p_1}^2 +2(1-\eta^2){\bf p_3p_1}-\eta^2 {\bf p_2}^2-2\eta {\bf p_4} \right)  \\
{\bf J}_{\begin{ytableau} \ & \ \\ \ & \ \end{ytableau}} &= \frac{3\eta^2}{(2\eta^2+1)(\eta^2+2)(\eta^2+1)^2} \left(\eta^2 {\bf p_1}^4+2(\eta-\eta^3){\bf p_2p_1}^2 - 4\eta^2 {\bf p_3p_1}+(\eta^4+\eta^2+1){\bf p_2}^2
+ (\eta^3-\eta) {\bf p_4}\right) \\
{\bf J}_{\begin{ytableau} \ & \ \\ \ \\ \ \end{ytableau}} &= \frac{2\eta}{(3\eta^2+1)(\eta^2+1)^2} \left(\eta {\bf p_1}^4+(1-3\eta^2){\bf p_2p_1}^2+2(\eta^3-\eta){\bf p_3p_1} -\eta {\bf p_2}^2
+ 2\eta^2 {\bf p_4}\right)  \\
{\bf J}_{\begin{ytableau} \ \\ \ \\ \ \\ \ \end{ytableau}} &= \frac{1}
{(3\eta^2+1)(2\eta^2+1)(\eta^2+1)} \left( {\bf p_1}^4-6\eta {\bf p_2p_1}^2+3\eta^2 {\bf p_2}^2+8\eta^2 {\bf p_3p_1}-6\eta^3 {\bf p_4}\right)
\end{aligned}
\end{align}

These polynomials are eigenfunctions of $\hat W_2$:
\begin{align}
\begin{aligned}
\hat W_2 \cdot {\bf J}_{\begin{ytableau} \ & \ & \ & \ \end{ytableau}} &= (\omega_{4,1}+\omega_{3,1} + \omega_{2,1}){\bf J}_{\begin{ytableau} \ & \ & \ & \ \end{ytableau}}  \\
\hat W_2 \cdot{\bf J}_{\begin{ytableau} \ & \ & \ \\ \ \end{ytableau}} &= (\omega_{3,1}+\omega_{2,1} + \omega_{1,2}){\bf J}_{\begin{ytableau} \ & \ & \ \\ \ \end{ytableau}} \\
\hat W_2 \cdot{\bf J}_{\begin{ytableau} \ & \ \\ \ & \ \end{ytableau}} &= (\omega_{2,2}+\omega_{2,1} + \omega_{1,2}){\bf J}_{\begin{ytableau} \ & \ \\ \ & \ \end{ytableau}}  \\
\hat W_2 \cdot{\bf J}_{\begin{ytableau} \ & \ \\ \ \\ \ \end{ytableau}} &= (\omega_{1,3}+\omega_{2,1} + \omega_{1,2}){\bf J}_{\begin{ytableau} \ & \ \\ \ \\ \ \end{ytableau}}  \\
\hat W_2 \cdot{\bf J}_{\begin{ytableau} \ \\ \ \\ \ \\ \ \end{ytableau}} &= (\omega_{1,4}+\omega_{1,3} + \omega_{1,2}){\bf J}_{\begin{ytableau} \ \\ \ \\ \ \\ \ \end{ytableau}}
\end{aligned}
\end{align}

Most of these facts about Jack polynomials are well known -- from other approaches,
like the one reviewed in \cite{Ker2Mironov:2018nie}.
The main result of this section is that our {\it current} method to define them
continues to work pretty well when
the number of Young tableaux exceeds the number of equations
(which, in its turn, continues to exceed the number of Young diagrams).
This encourages its application in the {\it terra incognita} --
to the study of $3$-Schurs, in a way which slightly deviates from Wang's.

\section{Towards 3-Schur polynomials}

Now we apply the procedure to the case of 3-Schur polynomials, taking into account appropriate modifications: 2D Young diagrams become 3D Young diagrams/plane partitions, cut-and-join operator $\hat W_2$ and the content-function $\omega$ are deformed.

Let the content-function have the form
\begin{equation}
    \omega_{i,j,k}:=\H_1 \cdot(i-1)+\H_2 \cdot (j-1) + \H_3 \cdot (k-1)
\end{equation}
where three parameters $\omega_1, \omega_2, \omega_3$ are bound by one condition $\sigma_1:=\H_1+\H_2+\H_3=0$. The other two invariant combinations $\sigma_2:=\H_1\H_2+\H_1\H_3+\H_2\H_3$,
$\sigma_3:=\H_1\H_2\H_3$ will enter the deformed cut-and-join operator. The case of Jack polynomials in the above sections corresponds to  $\H_1=-\eta, \H_2=\frac{1}{\eta}, \H_3=\eta-\frac{1}{\eta}$
and is distinguished by the vanishing of the product
$(1+\H_1\H_2)(1+\H_1\H_3)(1+\H_2\H_3)$
which eliminates the "extra" $3d$ polynomials.

Let the deformed cut-and-join operator have the following form
\begin{align}
\boxed{\boxed{
\begin{aligned}
\hat W_2 = \frac{1}{2}\sum_{a,b=1}^{\infty}\sum_{i=1}^a\sum_{j=1}^b
\left(A_{i+j-1} \cdot a \, b \cdot {\bf P}_{a + b, i + j - 1}\frac{\p^2}{\p {\bf P}_{a, i}\p {\bf P}_{b, j}}
+ B_{i+j-1} \cdot (a + b)\cdot{\bf P}_{a, i}{\bf P}_{b, j}\frac{\p}{\p {\bf P}_{a + b, i + j - 1}}\right)
+  \\
+ \frac{1}{2} \sum_{a=1}^{\infty} \sum_{i=1}^a u_{a,i} \, a \, (a - 1){\bf P}_{a, i}\frac{\p}{\p {\bf P}_{a, i}}
+ \frac{1}{2}\sum_{a=1}^{\infty} \sum_{i=1}^{a-1} a \, (a-1) \left( v_{a,i} \,  {\bf P}_{a, i}\frac{\p}{\p {\bf P}_{a, i+1}}
+ w_{a,i} \, {\bf P}_{a, i + 1}\frac{\p}{\p {\bf P}_{a, i}} \right)
\label{3SchursW2}
\end{aligned}
}}
\end{align}
with the triangular set of times ${\bf P}_{a,i}$, where $i \leqslant a$ and time-independent functions $A,B,u,v,w$. There are several comments on the form of the cut-and-join operator. The above ansatz is inspired by the form of 2d cut-and-join operator \eqref{cut and join Jack}. One can write the following general operators
\begin{equation}
\label{operators ansatz}
     {\bf P}_{a + b, k}\frac{\p^2}{\p {\bf P}_{a, i}\p {\bf P}_{b, j}}, \hspace{5mm} {\bf P}_{a, i}{\bf P}_{b, j}\frac{\p}{\p {\bf P}_{a + b, k}}, \hspace{5mm}   {\bf P}_{a, i}\frac{\p}{\p {\bf P}_{a, j}}
\end{equation}
that generalize 2d cut-and-join operator for the case when times have two indices. For first and second operators  in \eqref{operators ansatz} we consider only $k= i+j-1$ as the most simplest law applicable to 2d case (indeed, for 2d case all the times have the form ${\bf P}_{a,1}$ and $i=j=k=1$). For the third operator in \eqref{operators ansatz} we also consider the simplest opportunity $i=j\pm1$ and $i=j$. More involved form of ansatz for $\hat W_2$ will be considered in future works.

To reproduce the Wang's answers \cite{Wang1} we impose a normalization condition
\be
A_1B_1=1
\ee
This condition actually breaks the homogeneity and we discuss the other normalization in Section \ref{homogeneity}.
Define
\begin{align}
\hat e_0 & :={\bf P}_{1,1} \nn \\
\hat e_1 & :=[\hat W_2,\hat e_0] = \sum_{a=1}^{\infty}\sum_{i=1}^a A_i\cdot a \cdot {\bf P}_{a+1,i}\frac{\p}{\p {\bf P}_{a,i}}  \nn \\
\hat e_2 & :=[\hat W_2,\hat e_1] 
\end{align}
The factor $a-1$ in the last terms in (\ref{3SchursW2}) does not allow this term to contribute to $\hat e_1$,
making $\hat e_1$ a pure differential operator. 

Now the slightly modify the method from section \ref{Explanation of the method} in order to apply it to the more involved case of 3-Schur functions:
\begin{enumerate}
    \item We consider one more operator $\hat e_2$ to produce "words" at the l.h.s. of the system. It looks natural as we proceed from 2d, where we need 2 operators, to 3d case. The p.4 of \ref{Explanation of the method} is modified:
    
    We look at the positions $k$ of $\textcolor{red}{\hat e_i}$ in the "word" -- counted from the right.
    Then we look what is the box $\textcolor{red}{\Box_k}$ in the diagram labeled by this number and put the corresponding factor $\textcolor{red}{(\omega_{\Box_k})^{i}}$ in front of the factor  $C_{\bar R}$. 
    
    In other words, for $\hat e_2$ the factor is $(\omega_{\Box_k})^{2}$. This rule is inspired by the constructions of \cite{Prochazka:2015deb}.
    \item In 3d case we have additional unknown parameters - the coefficients of the ansatz \eqref{3SchursW2}, therefore we use the eigenfunction property as additional equation to solve:
    \be
        \hat W_2\, {\bf S}_R =   \lambda_R {\bf S}_R, \ \ \ \ \ \ \ \ \ \
        \lambda_R =  \sum_{\Box \in R} \omega_\Box
    \ee
\end{enumerate}

\subsection{Level 1}

At this level there is only one Young diagram and everything is trivial:
\begin{equation}
    \ytableausetup{boxsize = 0.5em}
{\bf S_{\begin{ytableau} \ \end{ytableau}}}:=\hat e_0\cdot 1 = {\bf P}_{1,1},
 \hspace{35mm} \hat W_2 \cdot {\bf S_{\begin{ytableau} \ \end{ytableau}}} = 0.
\end{equation}

\subsection{Level 2}
At this level there are three Young diagrams and corresponding 3-Schur polynomials:
\begin{align}
\begin{aligned}
\hat e_0 \hat e_0 \cdot 1 = {\bf P}_{1,1}^2 &= {\bf S_{\begin{ytableau} \ \\ \ \end{ytableau}}} +{\bf S_{\begin{ytableau} \ & \ \end{ytableau}}} + {\bf S_{\begin{ytableau} \ \end{ytableau},\begin{ytableau} \ \end{ytableau}}} \nn \\
\hat e_1 \hat e_0  \cdot 1 = A_1 {\bf P}_{2,1}
&= \omega_{2,1,1}{\bf S_{\begin{ytableau} \ \\ \ \end{ytableau}}} +\omega_{1,2,1}{\bf S_{\begin{ytableau} \ & \ \end{ytableau}}} + \omega_{1,1,2} {\bf S_{\begin{ytableau} \ \end{ytableau},\begin{ytableau} \ \end{ytableau}}} \nn \\
\hat e_2 \hat e_0 \cdot 1 = {\bf P}_{1,1}^2 + A_1 \sigma_3 {\bf P}_{2,1} +A_1 w_{2,1}{\bf P}_{2,2}
&=\omega_{2,1,1}^2{\bf S_{\begin{ytableau} \ \\ \ \end{ytableau}}} +\omega_{1,2,1}^2{\bf S_{\begin{ytableau} \ & \ \end{ytableau}}} + \omega_{1,1,2}^2 {\bf S_{\begin{ytableau} \ \end{ytableau},\begin{ytableau} \ \end{ytableau}}}
\end{aligned}
\end{align}
We omit constants $C_{\begin{ytableau} \yt{1} \\ \yt{2} \end{ytableau}} = C_{\begin{ytableau} \yt{1} & \yt{2} \end{ytableau}} = C_{\begin{ytableau} \yt{1} \end{ytableau},\begin{ytableau} \yt{2} \end{ytableau}} = 1$.
Then substitute into
\be
\hat W_2 \cdot {\bf S_{\begin{ytableau} \ \\ \ \end{ytableau}}} = \omega_{2,1,1}{\bf S_{\begin{ytableau} \ \\ \ \end{ytableau}}}, \ \ \ \ \hat W_2 \cdot {\bf S_{\begin{ytableau} \ & \ \end{ytableau}}} = \omega_{1,2,1}{\bf S_{\begin{ytableau} \ & \ \end{ytableau}}}, \ \ \ \
\hat W_2 \cdot {\bf S_{\begin{ytableau} \ \end{ytableau},\begin{ytableau} \ \end{ytableau}}} = \omega_{1,1,2}{\bf S_{\begin{ytableau} \ \end{ytableau},\begin{ytableau} \ \end{ytableau}}}
\ee
The above equations fix three parameters:
\be
u_{2,1} = \sigma_3, \ \ \ \
u_{2,2} = -\sigma_3, \ \ \ \
v_{2,1} = -\frac{(1+\H_1\H_2)(1+\H_1\H_3)(1+\H_2\H_3)}{ w_{2,1}}
\label{u v 2 level B = 1}
\ee
Then 
\begin{align}
    \boxed{
{\bf S_{\begin{ytableau} \ \\ \ \end{ytableau}}} =  \frac{(1+\H_2\H_3){ {\bf P}_{1,1}^2}  +A_1\H_1(1+\H_2\H_3) { {\bf P}_{2,1}}
+ w_{2,1} { {\bf P}_{2,2}}}
{(\H_1-\H_2)(\H_1-\H_3)}
}
\label{3-Schur 2 level B=1}
\end{align}
and the other two polynomials are obtained by permutations of $\H_i$. In other words, a permutation of $\H_i$ corresponds to transposition of the Young diagram in the corresponding plane:
\begin{equation}
    {\bf S_{\begin{ytableau} \ & \ \end{ytableau}}} = {\bf S_{\begin{ytableau} \ \\ \ \end{ytableau}}} \left( \H_1 \leftrightarrow \H_2 \right), \hspace{15mm} {\bf S_{\begin{ytableau} \ \end{ytableau},\begin{ytableau} \ \end{ytableau}}} = {\bf S_{\begin{ytableau} \ \\ \ \end{ytableau}}} \left( \H_1 \leftrightarrow \H_3 \right)
\end{equation}
In \cite{Wang1} the parameters are taken to be  $A_1=1$ and
$w_{2,1}=\sqrt{(1+\H_1\H_2)(1+\H_1\H_3)(1+\H_2\H_3)}$,
but in our calculation they are not yet fixed at this level.
Still we impose these constraints to simplify the formulas
and their reduction to the Jack case.

\subsection{Level 3}

Now we define the 6 polynomials ${\bf S_{\begin{ytableau} \ \\ \ \\ \ \end{ytableau}}},{\bf S_{\begin{ytableau} \ & \ & \ \end{ytableau}}},{\bf S_{\begin{ytableau} \ \end{ytableau},\begin{ytableau} \ \end{ytableau},\begin{ytableau} \ \end{ytableau}}},  {\bf S_{\begin{ytableau} \ & \ \\ \ \end{ytableau}}},  {\bf S_{\begin{ytableau} \ & \ \end{ytableau},\begin{ytableau} \ \end{ytableau}}}, {\bf S_{\begin{ytableau} \ \\ \ \end{ytableau},\begin{ytableau} \ \end{ytableau}}}$ from a system of 9 equations ($i,j = 0,1,2$):
\begin{align}
    \begin{aligned}
    \textcolor{red}{\hat e_i} \textcolor{blue}{\hat e_j} \hat e_0 \cdot 1 &= 
    \textcolor{red}{\left( \omega_{2,1,1} \right)^{i}} \textcolor{blue}{\left( \omega_{3,1,1} \right)^{j}} \,  {\bf S_{\begin{ytableau} \ \\ \ \\ \ \end{ytableau}}} +
    \textcolor{red}{\left( \omega_{1,2,1} \right)^{i}} \textcolor{blue}{\left( \omega_{1,3,1} \right)^{j}} \, {\bf S_{\begin{ytableau} \ & \ & \ \end{ytableau}}} +
    \textcolor{red}{\left( \omega_{1,1,2} \right)^{i}} \textcolor{blue}{\left( \omega_{1,1,3} \right)^{j}} \, {\bf S_{\begin{ytableau} \ \end{ytableau},\begin{ytableau} \ \end{ytableau},\begin{ytableau} \ \end{ytableau}}} + \\
    &+  \left( \textcolor{red}{\left( \omega_{2,1,1} \right)^{i}} \textcolor{blue}{\left( \omega_{1,2,1} \right)^{j}} C_{\begin{ytableau} \yt{1} & \textcolor{blue}{\yt{2}} \\ \textcolor{red}{\yt{3}} \end{ytableau}} + 
    \textcolor{red}{\left( \omega_{1,2,1} \right)^{i}} \textcolor{blue}{\left( \omega_{2,1,1} \right)^{j}} C_{\begin{ytableau} \yt{1} & \textcolor{red}{\yt{3}} \\ \textcolor{blue}{\yt{2}} \end{ytableau}} \right) {\bf S_{\begin{ytableau} \ & \ \\ \ \end{ytableau}}} + \\
    &+  \left( \textcolor{red}{\left( \omega_{1,1,2} \right)^{i}} \textcolor{blue}{\left( \omega_{1,2,1} \right)^{j}} C_{\begin{ytableau} \yt{1} & \textcolor{blue}{\yt{2}} \end{ytableau},\begin{ytableau} \textcolor{red}{\yt{3}} \end{ytableau}} + 
    \textcolor{red}{\left( \omega_{1,2,1} \right)^{i}} \textcolor{blue}{\left( \omega_{1,1,2} \right)^{j}} C_{\begin{ytableau} \yt{1} & \textcolor{red}{\yt{3}} \end{ytableau},\begin{ytableau} \textcolor{blue}{\yt{2}} \end{ytableau}} \right) {\bf S_{\begin{ytableau} \ & \ \end{ytableau},\begin{ytableau} \ \end{ytableau}}} + \\
    &+  \left( \textcolor{red}{\left( \omega_{1,1,2} \right)^{i}} \textcolor{blue}{\left( \omega_{2,1,1} \right)^{j}} C_{\begin{ytableau} \yt{1} \\ \textcolor{blue}{\yt{2}} \end{ytableau},\begin{ytableau} \textcolor{red}{\yt{3}} \end{ytableau}} + 
    \textcolor{red}{\left( \omega_{2,1,1} \right)^{i}} \textcolor{blue}{\left( \omega_{1,1,2} \right)^{j}} C_{\begin{ytableau} \yt{1} \\ \textcolor{red}{\yt{3}} \end{ytableau},\begin{ytableau} \textcolor{blue}{\yt{2}} \end{ytableau}} \right) {\bf S_{\begin{ytableau} \ \\ \ \end{ytableau},\begin{ytableau} \ \end{ytableau}}} 
    \end{aligned}
\end{align}
with the following condition:
\begin{equation}
    C_{\begin{ytableau} \yt{1} & \yt{2} \\ \yt{3} \end{ytableau}} + C_{\begin{ytableau} \yt{1} & \yt{3} \\ \yt{2} \end{ytableau}}  =  C_{\begin{ytableau} \yt{1} & \yt{2} \end{ytableau},\begin{ytableau} \yt{3} \end{ytableau}} + C_{\begin{ytableau} \yt{1} & \yt{3} \end{ytableau},\begin{ytableau} \yt{2} \end{ytableau}} = C_{\begin{ytableau} \yt{1} \\ \yt{2} \end{ytableau},\begin{ytableau} \yt{3} \end{ytableau}} + C_{\begin{ytableau} \yt{1} \\ \yt{3} \end{ytableau},\begin{ytableau} \yt{2} \end{ytableau}} = 2
    \label{sumrules}
\end{equation}
The rule to construct this equation is just the same as p.5 in section \ref{Explanation of the method}. As we mentioned above, the addition compared to 2d case is that each $\hat e_2$ at position  $k$ contributes a square of $\omega_{\Box_k}$.
The system is overdefined, but it is resolvable for appropriate choice of parameters.
Namely, we need 
\begin{align}
    \begin{aligned}
C_{\begin{ytableau} \yt{1} & \yt{3} \end{ytableau},\begin{ytableau} \yt{2} \end{ytableau}} &= \frac{2(2\H_2-\H_3)}{3(\H_2-\H_3)}, &\hspace{5mm}
C_{\begin{ytableau} \yt{1} \\ \yt{3} \end{ytableau},\begin{ytableau} \yt{2} \end{ytableau}} &= \frac{2(2\H_1-\H_3)}{3(\H_1-\H_3)}, &\hspace{5mm} C_{\begin{ytableau} \yt{1} & \yt{2} \\ \yt{3} \end{ytableau}} &= \frac{2(2\H_1-\H_2)}{3(\H_1-\H_2)}, \\
C_{\begin{ytableau} \yt{1} & \yt{2} \end{ytableau},\begin{ytableau} \yt{3} \end{ytableau}} &= \frac{2(2\H_3-\H_2)}{3(\H_3-\H_2)}, &\hspace{5mm}
C_{\begin{ytableau} \yt{1} \\ \yt{2} \end{ytableau},\begin{ytableau} \yt{3} \end{ytableau}} &= \frac{2(2\H_3-\H_1)}{3(\H_3-\H_1)},  &\hspace{5mm} C_{\begin{ytableau} \yt{1} & \yt{3} \\ \yt{2} \end{ytableau}} &= \frac{2(2\H_2-\H_1)}{3(\H_2-\H_1)}
    \end{aligned}
\end{align}
and also
\be
u_{3,1}=\sigma_3,  \hspace{20mm}   w_{3,1} = \frac{A_2}{A_1}w_{2,1}
\ee

Then we can ask if the polynomials ${\bf S}_{\pi}$, deduced from this resolvable system can be
eigenfunctions of $\hat W_2$.
This is not just automatic, 
the eigenfunction conditions
\begin{align}
\begin{aligned}
\hat W_2 \cdot {\bf S_{\begin{ytableau} \ \\ \ \\ \ \end{ytableau}}} &= (\omega_{3,1,1}+\omega_{2,1,1}){\bf S_{\begin{ytableau} \ \\ \ \\ \ \end{ytableau}}} \\
\hat W_2 \cdot {\bf S_{\begin{ytableau} \ & \ & \ \end{ytableau}}} &= (\omega_{1,3,1}+\omega_{1,2,1}) {\bf S_{\begin{ytableau} \ & \ & \ \end{ytableau}}}  \\
\hat W_2 \cdot {\bf S_{\begin{ytableau} \ \end{ytableau},\begin{ytableau} \ \end{ytableau},\begin{ytableau} \ \end{ytableau}}} &= (\omega_{1,1,3}+\omega_{1,1,2}){\bf S_{\begin{ytableau} \ \end{ytableau},\begin{ytableau} \ \end{ytableau},\begin{ytableau} \ \end{ytableau}}}  \\
\hat W_2 \cdot {\bf S_{\begin{ytableau} \ & \ \end{ytableau},\begin{ytableau} \ \end{ytableau}}} &= (\omega_{1,2,1}+\omega_{1,1,2}){\bf S_{\begin{ytableau} \ & \ \end{ytableau},\begin{ytableau} \ \end{ytableau}}} \\
\hat W_2 \cdot {\bf S_{\begin{ytableau} \ \\ \ \end{ytableau},\begin{ytableau} \ \end{ytableau}}} &= (\omega_{2,1,1}+\omega_{1,1,2}){\bf S_{\begin{ytableau} \ \\ \ \end{ytableau},\begin{ytableau} \ \end{ytableau}}}  \\
\hat W_2 \cdot {\bf S_{\begin{ytableau} \ & \ \\ \ \end{ytableau}}} &= (\omega_{2,1,1}+\omega_{1,2,1}){\bf S_{\begin{ytableau} \ & \ \\ \ \end{ytableau}}} \\
\end{aligned}
\end{align}
are satisfied, provided {\it either}
\be
A_2B_2=-\frac{1}{3}, \ \ \ \  u_{3,2}=-\sigma_3, \ \ \ u_{3,3}=0, \ \ \
v_{3,1} = -\frac{2A_1}{A_2} \frac{(1+\H_1\H_2)(1+\H_1\H_3)(1+\H_2\H_3)}{w_{2,1}}, \ \ \ \
v_{3,2} =  \frac{2}{3w_{3,2}}
\label{cho31}
\ee
or
\be
A_2B_2=\frac{2}{3}, \ \ \ \  u_{3,2}=u_{3,3}=-\frac{\sigma_3}{2}, \ \ \
v_{3,1} = -\frac{2A_1}{3A_2} \frac{(1+\H_1\H_2)(1+\H_1\H_3)(1+\H_2\H_3)}{w_{2,1}}, \nn \\
v_{3,2} = - \frac{(2+\H_1\H_2)(2+\H_1\H_3)(2+\H_2\H_3)}{12w_{3,2}}
\label{cho32}
\ee
In what follows we make the second choice (\ref{cho32}) as it leads to Wang's answer. The nature of the other solution \eqref{cho31} remains unclear. 
Then 
\begin{align}
\boxed{
\begin{aligned}
{\bf S_{\begin{ytableau} \ \\ \ \\ \ \end{ytableau}}} &= 
\frac{1}
{(\H_1-\H_2)(\H_1-\H_3)(2\H_1-\H_2)(2\H_1-\H_3)}
\Big[ (1+\H_2\H_3)(2+\H_2\H_3){\bf P}_{1,1}^3 + \\
& + 3A_1\H_1(1+\H_2\H_3)(2+\H_2\H_3){\bf P}_{2,1}{\bf P}_{1,1} + 3A_1 (2+\H_2\H_3)w_{2,1} {\bf P}_{2,2}{\bf P}_{1,1} +  \\
& +2A_1^2\H_1^2(1+\H_2\H_3)(2+\H_2\H_3){\bf P}_{3,1}  +3A_1A_2w_{2,1}\H_1(2+\H_2\H_3) {\bf P}_{3,2} 
+ 6A_1A_2w_{2,1}w_{3,2}{\bf P}_{3,3}\Big]
\\ \\ 
{\bf S_{\begin{ytableau} \ & \ \\ \ \end{ytableau}}} &= 
\frac{3}{(\H_3-\H_1)(\H_3-\H_2)(2\H_1-\H_2)(2\H_2-\H_1)}
\Big[ (1+\H_2\H_3)(1+\H_1\H_3){\bf P}_{1,1}^3  
+  \\
 &- A_1 \H_3 (1+\H_2\H_3)(1+\H_1\H_3){\bf P}_{2,1}{\bf P}_{1,1}
+ A_1 (3-\H_3^2)w_{2,1} {\bf P}_{2,2}{\bf P}_{1,1}
+   \\
&+A_1^2\H_1\H_2(1+\H_2\H_3)(1+\H_1\H_3){\bf P}_{3,1}
 +A_1A_2w_{2,1}(-\H_3 + 3/2\H_1\H_2\H_3) {\bf P}_{3,2} 
 + 3A_1A_2w_{2,1}w_{3,2}{\bf P}_{3,3}
\Big]
\end{aligned}
}
\end{align}

For an appropriate choice of constants:
\begin{align}
    \begin{aligned}
    A_1 &= 1, &\hspace{15mm} w_{2,1} &= \sqrt{(1+\H_1\H_2)(1+\H_1\H_3)(1+\H_2\H_3)}, \\
    A_2 &=1, &\hspace{15mm} w_{3,2} &= \frac{1}{6} \sqrt{(2+\H_1\H_2)(2+\H_1\H_3)(2+\H_2\H_3)}. 
    \end{aligned}
\end{align}
this reproduces the result of \cite{Wang1}.

We do not present the results and implications of the complicated analysis
of the level-4 example in this short paper.
They will be described in a separate more technical presentation.

\section{On the definition/evaluation of $C_{\bar \pi}$}
\label{C consts}
The crucial ingredient of the Yangian approach to generalized Schur polynomials
(like Jacks for Young diagrams and $3$-Schurs for plane partitions)
is the idea that they are associated with {\it tableaux} rather than with
{\it diagrams}, i.e. remember the sequence of box additions which lead to
building up the diagram.
Thus we get a huge set of polynomials ${\cal S}_{\bar \pi}$ at the r.h.s. of our defining relations
\be
{\rm words}\{\hat e_2, \hat e_1,\hat e_0\}\cdot 1 = {\cal G}\{\omega\}\cdot  \{{\cal S}_{\hat \pi}\}
\label{Gmat}
\ee
${\cal G}\{\omega\}$ here is the known  matrix, build from the  function
$\omega_{\Box} = \omega_{i,j,k} = \H_1 \cdot (i-1)+\H_2 \cdot (j-1) + \H_3 \cdot (k-1)$,
but  it is rectangular at each level $l$, of the size
$d^{l-1}\times {\rm tab}_d^{(l)}$, where $d=3$ and ${\rm tab}_d^{(l)}$ is the number
of tables.
The problem is that while $d^{l-1}$ exceeds the number of {\it diagrams},
it is smaller than the number of {\it tables}:
\be
{\rm dia}_d^{(l)}\leq d^{l-1}  \leq {\rm tab}_d^{(l)}
\ee
This makes the matrix rectangular in the wrong way -- the system of equations for $P_{\bar\pi}$
is {\it under}defined and does not have a unique solution.

There are three ways to handle this problem.

{\bf First}, one can extend the number of words, considering the action of higher operators
$\hat e_a$, generated by the same recursion rule $\hat e_{a}=[\hat W_2,\hat e_{a-1}]$.
This would add more lines to the matrix ${\cal G}$ (made from powers $\omega_{i,j,k}^a$),
and make the system {\it over}defined.
Still it will have solutions -- due to conspiracy, implied by Yangian representation theory.
This approach is most straightforward in the approach of this paper,
but it is computationally very hard, because it deals with a linear problem for
very big matrices.
The above-mentioned conspiracy allows to drastically simplify the calculation --
moreover, at least two additional steps are available, which we describe as
the two next options in our list.

{\bf Second}, we can use the fact that the dependence of $P_{\bar \pi}$ on the {\it table} is minor:
\be
{\cal S}_{\bar \pi}\{{\bf P}\} = C_{\bar \pi} {\bf S}_{\pi}\{{\bf P}\}
\ee
where $C_{\bar\pi}$ are {\it constants}, i.e. do not depend on the time-variables ${\bf P}$.
This substitutes the system (\ref{Gmat})
\be
\boxed{
{\rm words}\{\hat e_2, \hat e_1,\hat e_0\}\cdot 1 = { G}\{\omega\}\cdot    \{{\bf S}_{ \pi}\}
}
\label{Gmatred}
\ee
with a much smaller $d^{l-1}\times {\rm dia}_d^{(l)}$ matrix $G\{\omega\}={\cal G}\{\omega\}\cdot C_{\bar \pi}$
which is already rectangular in the proper way -- the linear system for ${\bf S}_{\pi}$ is {\it over}defined.
Indeed, ${\rm dia}_d^{(l)}$ is the number of {\it diagrams},
generated by MacMahon function
\be
\sum_l {\rm dia}_3^{(l)}q^l = \prod_n \frac{1}{(1-q^n)^n}
= 1+q+3q^2+6q^3+13q^4+24q^5+48q^6 +86q^7 +160q^8 +282q^9+ \ldots
\nn \\
< 1+\sum_{l=1} 3^{l-1}q^l
\ee
This is the $3d$ analogue of the more familiar
\be
\sum_l {\rm dia}_2^{(l)}q^l = \prod_n \frac{1}{(1-q^n)}
= 1+q+2q^2+3q^3+5q^4+7q^5+11q^6 +15q^7 +22q^8 +30q^9+ \ldots
\nn \\
< 1+\sum_{l=1} 2^{l-1}q^l
\ee
To compare,
\be
\sum_l {\rm tab}_3^{(l)}q^l 
= 1+q+3q^2+9q^3+33q^4+135q^5+633q^6 +3207q^7 +17602q^8 +103041q^9+ \ldots
\nn
\ee
\vspace{-0.4cm}
\be
\hspace{11cm}    > 1+\sum_{l=1} 3^{l-1}q^l
\ee
and
\be
\sum_l {\rm tab}_2^{(l)}q^l
= 1+q+2q^2+4q^3+10q^4+26q^5+76q^6 +232q^7 +764q^8 +2620q^9+ \ldots
\nn \\
> 1+\sum_{l=1} 2^{l-1}q^l
\ee

Solution to the system (\ref{Gmatred}) now depends on the choice of the constants $C_{\bar \pi}$,
but since the system is {\it over}defined, there are additional constraints, which actually allow
to fix these constants.
We illustrated this method in consideration of Jacks for $d=2$.
Still, at $d=3$ it is still computationally difficult: although the linear system for ${\bf S}_{\pi}$
is now much smaller, one needs to keep additional parameters $C_{\bar \pi}$ in its coefficients,
and there are a lot.

{\bf Third}, one can deduce the expressions of $C_{\bar \pi}$ through $\omega_{i,j,k}$
from representation theory, and construct matrix $G\{\omega\}$ in (\ref{Gmatred})
explicitly.
Then what remains is just to solve the linear system (\ref{Gmatred}) for ${\bf S}_{\pi}$.
This is the strategy which we actually follow for $3$-Schurs in this paper.

Explicit formulas for $C_{\bar \pi}$ are implied by the rule \cite{Tsymbaliuk:2014fvq,Prochazka:2015deb,Wang1}
to add boxes to the {\it table}:
\be
\label{C constant}
\tilde{ C}_{\bar\pi + \Box} = \tilde { C}_{\bar \pi}\cdot
\sqrt{\frac{\omega_\Box+\sigma_3}{\omega_\Box\sigma_3}\cdot
{\rm res}_\epsilon\!\! \prod_{\Box'\in \pi} \phi(\omega_\Box - \omega_{\Box'})}
\ee
where
\be
\phi(u):=\frac{(u+\H_1)(u+\H_2)(u+\H_3)}{(u-\H_1+\epsilon)(u-\H_2+\epsilon)(u-\H_3+\epsilon)}
\ee
The product over $\pi$ is actually singular at $\epsilon=0$ and one needs to take a residue
at the pole $\epsilon^{-1}$.
To get $\tilde C_{\bar \pi}$ one needs to take a  product of $|\pi|-1$ such square roots for
addition of every box in the process of building up $\bar\pi$.
We put tilde over $\tilde{C}_{\bar \pi}$, because these polynomials are actually normalized differently
from ours -- to get our normalization we impose conditions like (\ref{sumrules}) on $C_{\bar \pi}$.
To obtain our $C_{\bar \pi}$ we need the ratios of these ${\tilde C}_{\bar \pi}$, which are actually
full squares, i.e. the square root will not show up in the answers for $C_{\bar\pi}$.

There are simple and elegant formula for the ratios of ${\tilde C}_{\bar \pi}$ that explains the absence of square roots in the final answer. The formula was suggested to us by the anonymous reviewer:

\begin{equation}
\label{ratio of C}
    \frac{{\tilde C}_{\bar \pi}}{{\tilde C}_{\bar \pi^{\prime}}} = \prod_{\text{exchanges of boxes} \ \Box, \Box^{\prime}} \phi\left( \omega_{\Box} - \omega_{\Box^{\prime}} \right)
\end{equation}
where the product is over all pair exchanges of boxes needed to transform $\bar \pi$ to $\bar \pi^{\prime}$. The result is finite and no residues and square roots need to be taken. 

There is a comment in order to clarify the above formula. The boxes $\Box$ and $\Box^{\prime}$ should have adjacent numbers in the Young tableaux and the number of $\Box$ should be greater, otherwise the ratio is inversed. For example:
\begin{equation}
    \frac{\tilde {C}_{\begin{ytableau} \yt{1} & \yt{2} \\ \yt{3} \end{ytableau}}}{\tilde {C}_{\begin{ytableau} \yt{1} & \yt{3} \\ \yt{2} \end{ytableau}}} = \phi( \omega_{2,1,1} - \omega_{1,2,1} ), 
    \hspace{10mm} 
     \frac{\tilde {C}_{\begin{ytableau} \yt{1} & \yt{3} \\ \yt{2} \end{ytableau}}}{\tilde {C}_{\begin{ytableau} \yt{1} & \yt{2} \\ \yt{3} \end{ytableau}}} = \phi( \omega_{1,2,1} - \omega_{2,1,1} )
\end{equation}
Here the boxes with numbers $2,3$ are exchanged and the formula is correct. We provide another example to show that exchanging boxes with non adjacent numbers gives wrong result:
\begin{equation}
    \frac{\tilde{C}_{\begin{ytableau} \yt{1} & \yt{2} \\ \yt{4}\end{ytableau}, \begin{ytableau} \yt{3} \end{ytableau}}}{\tilde{C}_{\begin{ytableau} \yt{1} & \yt{4} \\ \yt{2}\end{ytableau}, \begin{ytableau} \yt{3} \end{ytableau}}}  \not= \phi(\omega_{2,1,1} -\omega_{1,2,1}) 
\end{equation}
Namely, one can not exchange boxes with numbers $2$ and $4$ via one transposition. One should do in three steps $4 \leftrightarrow 3$, $3 \leftrightarrow 2$ and then $4 \leftrightarrow 3$:
\begin{equation}
    \frac{\tilde{C}_{\begin{ytableau} \yt{1} & \yt{2} \\ \yt{4}\end{ytableau}, \begin{ytableau} \yt{3} \end{ytableau}}}{\tilde{C}_{\begin{ytableau} \yt{1} & \yt{4} \\ \yt{2}\end{ytableau}, \begin{ytableau} \yt{3} \end{ytableau}}} =  \phi(\omega_{2,1,1} -\omega_{1,1,2}) \cdot \phi(\omega_{2,1,1} -\omega_{1,2,1})  \cdot \phi(\omega_{1,1,2} -\omega_{1,2,1})
\end{equation}

Now we provide a few examples of calculation at levels three and four using normalization condition from p.5 section \ref{Explanation of the method}.

\subsection*{ Example. Level 3}

Here we have three pairs of interesting diagrams of which we consider just one, relevant for
${\bf S}_{\begin{ytableau} \ & \ \\ \ \end{ytableau}}$:
\begin{align}
\begin{aligned}
\tilde {C}_{\begin{ytableau} \yt{1} & \yt{2} \\ \yt{3} \end{ytableau}} &=
\sqrt{  \left[\frac{\omega_{2,1,1}+\sigma_3}{\omega_{2,1,1}\sigma_3} {\rm res}_\epsilon
\Big(\phi(\omega_{2,1,1}-\omega_{1,2,1})\phi(\omega_{2,1,1})\Big)\right]
\cdot \left[\frac{\omega_{1,2,1}+\sigma_3}{\omega_{1,2,1}\sigma_3} {\rm res}_\epsilon
\phi(\omega_{1,2,1})\right]
}  
\nn \\
\tilde {C}_{\begin{ytableau} \yt{1} & \yt{3} \\ \yt{2} \end{ytableau}} &=
\sqrt{  \left[\frac{\omega_{1,2,1}+\sigma_3}{\omega_{1,2,1}\sigma_3} {\rm res}_\epsilon
\Big(\phi(\omega_{1,2,1}-\omega_{2,1,1})\phi(\omega_{1,2,1})\Big)\right]
\cdot \left[\frac{\omega_{2,1,1}+\sigma_3}{\omega_{2,1,1}\sigma_3} {\rm res}_\epsilon
\phi(\omega_{2,1,1})\right]
} 
\end{aligned}
\end{align}
These constants $\tilde {C}_{\begin{ytableau} \yt{1} & \yt{3} \\ \yt{2} \end{ytableau}}, \tilde {C}_{\begin{ytableau} \yt{1} & \yt{2} \\ \yt{3} \end{ytableau}}$ differ from ours $C_{\begin{ytableau} \yt{1} & \yt{2} \\ \yt{3} \end{ytableau}}, C_{\begin{ytableau} \yt{1} & \yt{2} \\ \yt{3} \end{ytableau}}$ by overall factor:
\begin{equation}
    \tilde {C}_{\begin{ytableau} \yt{1} & \yt{3} \\ \yt{2} \end{ytableau}} + \tilde {C}_{\begin{ytableau} \yt{1} & \yt{2} \\ \yt{3} \end{ytableau}} \not = 2
\end{equation}

\begin{equation}
    C_{\begin{ytableau} \yt{1} & \yt{3} \\ \yt{2} \end{ytableau}} + C_{\begin{ytableau} \yt{1} & \yt{2} \\ \yt{3} \end{ytableau}} = 2,
\end{equation}
however, the ratio is the same:
\begin{equation}
     \frac{C_{\begin{ytableau} \yt{1} & \yt{2} \\ \yt{3} \end{ytableau}}}{C_{\begin{ytableau} \yt{1} & \yt{3} \\ \yt{2} \end{ytableau}}} = \frac{\tilde {C}_{\begin{ytableau} \yt{1} & \yt{2} \\ \yt{3} \end{ytableau}}}{\tilde {C}_{\begin{ytableau} \yt{1} & \yt{3} \\ \yt{2} \end{ytableau}}} =  \phi( \omega_{1,2,1} - \omega_{2,1,1} ) = \frac{2\H_1-\H_2}{\H_1-2\H_2}
\end{equation}
These equations are enough to compute coefficients in our normalization. In this particular case we reproduce the result of previous sections:
\begin{equation}
    C_{\begin{ytableau} \yt{1} & \yt{2} \\ \yt{3} \end{ytableau}} = \frac{2(2\H_1-\H_2)}{3(\H_1-\H_2)}, \hspace{15mm} C_{\begin{ytableau} \yt{1} & \yt{3} \\ \yt{2} \end{ytableau}} = \frac{2(2\H_2-\H_1)}{3(\H_2-\H_1)}
\end{equation}
Substituting $\H_1=-\eta$, $\H_2=\eta^{-1}$ we get the answer for the Jack case \eqref{3 level Jacks}:
\begin{equation}
    C_{\begin{ytableau} \yt{1} & \yt{2} \\ \yt{3} \\ \end{ytableau}} =\frac{2(2\eta^2+1)}{3(\eta^2+1)}, \hspace{15mm} C_{\begin{ytableau} \yt{1} & \yt{3} \\ \yt{2} \\ \end{ytableau}} = \frac{2(\eta^2+2)}{3(\eta^2+1)}
\end{equation}

\subsection*{ Example. Level 4}

Using elegant formula \eqref{ratio of C} we obtain:
\be
C_{\begin{ytableau} \yt{1} & \yt{3} \\ \yt{2} \\ \yt{4} \end{ytableau}}
= \frac{(3\H_1-\H_2)(\H_1-2\H_2)}{3(\H_1-\H_2)^2}\cdot
C_{\begin{ytableau} \yt{1} & \yt{4} \\ \yt{2} \\ \yt{3} \end{ytableau}},
\nn \\
C_{\begin{ytableau} \yt{1} & \yt{2} \\ \yt{3} \\ \yt{4} \end{ytableau}}
= \frac{(3\H_1-\H_2)(2\H_1-\H_2)}{3(\H_1-\H_2)^2}
\cdot C_{\begin{ytableau} \yt{1} & \yt{4} \\ \yt{2} \\ \yt{3} \end{ytableau}} 
\ee
Using normalization condition $C_{\begin{ytableau} \yt{1} & \yt{2} \\ \yt{3} \\ \yt{4} \end{ytableau}} + C_{\begin{ytableau} \yt{1} & \yt{3} \\ \yt{2} \\ \yt{4} \end{ytableau}} +
C_{\begin{ytableau} \yt{1} & \yt{4} \\ \yt{2} \\ \yt{3} \end{ytableau}} = 3$ we compute the constants:
\be
C_{\begin{ytableau} \yt{1} & \yt{4} \\ \yt{2} \\ \yt{3} \end{ytableau}} = \frac{3(\H_1-\H_2)}{2(2\H_1-\H_2)},
\hspace{5mm}
C_{\begin{ytableau} \yt{1} & \yt{3} \\ \yt{2} \\ \yt{4} \end{ytableau}}
= \frac{(3\H_1-\H_2)(\H_1-2\H_2)}{2(2\H_1-\H_2)(\H_1-\H_2)},
\hspace{5mm}
C_{\begin{ytableau} \yt{1} & \yt{2} \\ \yt{3} \\ \yt{4} \end{ytableau}} = \frac{3\H_1-\H_2}{2(\H_1-\H_2)}
\label{12C31}
\ee
what reproduce the Jack values \eqref{C constant level 4} at $\H_1=-\eta$, $\H_2=\eta^{-1}$:
\be
C_{\begin{ytableau} \yt{1} & \yt{4} \\ \yt{2} \\ \yt{3} \end{ytableau}} = \frac{3(\eta^2+1)}{2(2\eta^2+1)},
\hspace{5mm}
C_{\begin{ytableau} \yt{1} & \yt{3} \\ \yt{2} \\ \yt{4} \end{ytableau}}
=\frac{(3\eta^2+1)(\eta^2+2)}{2(2\eta^2+1)(\eta^2+1)},
\hspace{5mm}
C_{\begin{ytableau} \yt{1} & \yt{2} \\ \yt{3} \\ \yt{4} \end{ytableau}} = \frac{3\eta^2+1}{2(\eta^2+1)}
\ee
Constants $C_{\bar \pi}$ for the other tables that lie in different planes are obtained by the permutations of $\H_1, \H_2, \H_3$
in (\ref{12C31}). For example, $\H_1 \leftrightarrow \H_2$:
\be
C_{\begin{ytableau} \yt{1} & \yt{2} & \yt{3} \\ \yt{4}\end{ytableau}} = \frac{3(\H_2-\H_1)}{2(2\H_2-\H_1)},
\hspace{5mm}
C_{\begin{ytableau} \yt{1} & \yt{2} & \yt{4} \\ \yt{3} \end{ytableau}}
= \frac{(3\H_2-\H_1)(\H_2-2\H_1)}{2(2\H_2-\H_1)(\H_2-\H_1)},
\hspace{5mm}
C_{\begin{ytableau} \yt{1} & \yt{3} & \yt{4} \\ \yt{2} \end{ytableau}} = \frac{3\H_2-\H_1}{2(\H_2-\H_1)}
\ee

Likewise we deduce
\be
C_{\begin{ytableau} \yt{1} & \yt{3} \\ \yt{2} & \yt{4} \end{ytableau}}
= \frac{\H_1-2\H_2}{2\H_1 - \H_2} \cdot C_{\begin{ytableau} \yt{1} & \yt{2} \\ \yt{3} & \yt{4} \end{ytableau}}
\ee
and
\be
C_{\begin{ytableau} \yt{1} & \yt{3} \\ \yt{2} & \yt{4} \end{ytableau}} = \frac{2(2\H_2-\H_1)}{3(\H_2-\H_1)}
\ \longrightarrow \ \frac{2(\eta^2+2)}{3(\eta^2+1)},
\ \ \ \ \ \ \
C_{\begin{ytableau} \yt{1} & \yt{2} \\ \yt{3} & \yt{4} \end{ytableau}}
= 2- C_{\begin{ytableau} \yt{1} & \yt{3} \\ \yt{2} & \yt{4} \end{ytableau}} = \frac{2(2\H_1-\H_2)}{3(\H_1-\H_2)}
\ \longrightarrow \ \frac{2(2\eta^2+1)}{3(\eta^2+1)}
\ee

Finally, using formula \eqref{ratio of C}
\begin{align}
\tilde{C}_{\begin{ytableau} \yt{1} & \yt{2} \\ \yt{4}\end{ytableau}, \begin{ytableau} \yt{3} \end{ytableau}}
&= \frac{\H_2-2\H_3}{2\H_2-\H_3} \cdot \tilde{C}_{\begin{ytableau} \yt{1} & \yt{3} \\ \yt{4}\end{ytableau}, \begin{ytableau} \yt{2} \end{ytableau}}
\nn \\ 
\tilde{C}_{\begin{ytableau} \yt{1} & \yt{4} \\ \yt{3}\end{ytableau}, \begin{ytableau} \yt{2} \end{ytableau}}
&=\frac{\H_1-2\H_2}{2\H_1-\H_2} \cdot \tilde{C}_{\begin{ytableau} \yt{1} & \yt{3} \\ \yt{4}\end{ytableau}, \begin{ytableau} \yt{2} \end{ytableau}}
\nn \\
\tilde{C}_{\begin{ytableau} \yt{1} & \yt{4} \\ \yt{2}\end{ytableau}, \begin{ytableau} \yt{3} \end{ytableau}}
&=\frac{(\H_1-2\H_3)(\H_1-2\H_2)}{(2\H_1-\H_3)(2\H_1-\H_2)} \cdot \tilde{C}_{\begin{ytableau} \yt{1} & \yt{3} \\ \yt{4}\end{ytableau}, \begin{ytableau} \yt{2} \end{ytableau}}
\nn \\
\tilde{C}_{\begin{ytableau} \yt{1} & \yt{3} \\ \yt{2}\end{ytableau}, \begin{ytableau} \yt{4} \end{ytableau}}
&= \frac{(\H_1-2\H_2)(\H_1-2\H_3)(\H_2-2\H_3)}{(2\H_1-\H_2)(2\H_1-\H_3)(2\H_2-\H_3)} \cdot \tilde{C}_{\begin{ytableau} \yt{1} & \yt{3} \\ \yt{4}\end{ytableau}, \begin{ytableau} \yt{2} \end{ytableau}}
\nn \\
\tilde{C}_{\begin{ytableau} \yt{1} & \yt{2} \\ \yt{3}\end{ytableau}, \begin{ytableau} \yt{4} \end{ytableau}}
&=\frac{(\H_1-2\H_3)(\H_2-2\H_3)}{(2\H_1-\H_3)(2\H_2-\H_3)} \cdot \tilde{C}_{\begin{ytableau} \yt{1} & \yt{3} \\ \yt{4}\end{ytableau}, \begin{ytableau} \yt{2} \end{ytableau}}
\end{align}
As usual, the ratios are nice functions without square roots.
Then
\be
\left(1 + \frac{\H_2-2\H_3}{2\H_2-\H_3} + \frac{\H_1-2\H_2}{2\H_1-\H_2}
+ \frac{(\H_1-2\H_3)(\H_1-2\H_2)}{(2\H_1-\H_3)(2\H_1-\H_2)} +
\ \ \ \ \ \ \ \ \ \ \ \ \ \ \ \ \ \ \ \ \ \ \ \ \ \ \ \ \ \ \ \ \ \ \ \ \ \ \ \  \right. \nn \\ \left.
+ \frac{(\H_1-2\H_2)(\H_1-2\H_3)(\H_2-2\H_3)}{(2\H_1-\H_2)(2\H_1-\H_3)(2\H_2-\H_3)}
+ \frac{(\H_1-2\H_3)(\H_2-2\H_3)}{(2\H_1-\H_3)(2\H_2-\H_3)}
\right)\cdot
C_{\begin{ytableau} \yt{1} & \yt{3} \\ \yt{4}\end{ytableau}, \begin{ytableau} \yt{2} \end{ytableau}} = 6
\ee
and
\be
C_{\begin{ytableau} \yt{1} & \yt{3} \\ \yt{4}\end{ytableau}, \begin{ytableau} \yt{2} \end{ytableau}} =
\frac{2(2\H_1-\H_2)(2\H_1-\H_3)(2\H_2-\H_3)}{7(\H_1-\H_2)(\H_1-\H_3)(\H_2-\H_3)},
\ \ \ \ \
C_{\begin{ytableau} \yt{1} & \yt{2} \\ \yt{4}\end{ytableau}, \begin{ytableau} \yt{3} \end{ytableau}} =
\frac{2(2\H_1-\H_2)(2\H_1-\H_3)(\H_2-2\H_3)}{7(\H_1-\H_2)(\H_1-\H_3)(\H_2-\H_3)}
\nn \\
C_{\begin{ytableau} \yt{1} & \yt{4} \\ \yt{3}\end{ytableau}, \begin{ytableau} \yt{2} \end{ytableau}} =
\frac{2(\H_1-2\H_2)(2\H_1-\H_3)(2\H_2-\H_3)}{7(\H_1-\H_2)(\H_1-\H_3)(\H_2-\H_3)},
\ \ \ \ \
C_{\begin{ytableau} \yt{1} & \yt{4} \\ \yt{2}\end{ytableau}, \begin{ytableau} \yt{3} \end{ytableau}} =
\frac{2(\H_1-2\H_2)(\H_1-2\H_3)(2\H_2-\H_3)}{7(\H_1-\H_2)(\H_1-\H_3)(\H_2-\H_3)}
\nn \\
\!\!\!\!\!
C_{\begin{ytableau} \yt{1} & \yt{3} \\ \yt{2}\end{ytableau}, \begin{ytableau} \yt{4} \end{ytableau}} =
\frac{2(\H_1-2\H_2)(\H_1-2\H_3)(\H_2-2\H_3)}{7(\H_1-\H_2)(\H_1-\H_3)(\H_2-\H_3)},
\ \ \ \ \
C_{\begin{ytableau} \yt{1} & \yt{2} \\ \yt{3}\end{ytableau}, \begin{ytableau} \yt{4} \end{ytableau}} =
\frac{2(2\H_1-\H_2)(\H_1-2\H_3)(\H_2-2\H_3)}{7(\H_1-\H_2)(\H_1-\H_3)(\H_2-\H_3)}
\ee
As one can see all these coefficients correctly transform into each other under permutations of $\H_1, \H_2, \H_3$ that corresponds to transpositions.
These constants remain non-vanishing in the Jack case --
what should vanishe  in this case, is the polynomial ${\bf S}_{\begin{ytableau} \ & \ \\ \ \end{ytableau}, \begin{ytableau} \ \end{ytableau}}$

\section{Restoration of homogeneity}
\label{homogeneity}

To summarize, we investigated the possibility to postulate the cut-and-join operator
in a rather simple form and  impose a normalization condition
\be
A_1B_1=\rho
\ee

After that we ask:
\begin{align}
    \begin{aligned}
    {\rm  word}\{\hat e_i\}\cdot 1   &= {\cal G} \cdot {\cal S} = G \cdot {\bf S}    \\
 \hat W_2 \cdot {\bf S}_{\pi} &= \sum_{\Box \in \pi} \omega_\Box \cdot {\bf S}_{\pi}
    \end{aligned}
\end{align}
The compatibility conditions for the linear system fix the form of unknown functions $A,B,u,v,w$ in our ansatz for cut-and-join operator $\hat W_2$. 

In the main text we just put $\rho=1$ to make the formulas looking more like those in \cite{Wang1}.
However, this breaks a natural grading in powers of $\H_i$ and times ${\bf P}_{a,i}$.
A better choice implies that $\rho$ has degree two, 
and things simplify a lot for particular choice:
\begin{equation}
\boxed{\boxed{
    A_1 = -\sigma_2 \hspace{5mm} B_1 = 1
}}
\end{equation}
For this choice at least the "classical" part of cut-and-join operator has degree 1, provided that the degree of times and $\H_{i}$ is equal to one:
\begin{equation}
    \text{deg}\left[ {\bf P}_{a,i} \right] =\text{deg}\left[ \H_i \right] = 1
\end{equation}

To illustrate this idea we provide an example on 2 level. The restoration of grading can be seen in the case of cut-and-join  operator $\hat W_2$
\be
\boxed{
u_{2,1}  = -\frac{\sigma_3}{\sigma_2}, \hspace{10mm}   u_{2,2} = \frac{\sigma_3}{\sigma_2}, \hspace{10mm}  v_{2,1} = -\frac{\sigma_3}{\sigma_2} \hspace{10mm} w_{2,1} = \frac{\sigma_3}{\sigma_2}
}
\ee
and 3-Schur polynomials:
\begin{equation}
\boxed{
    {\bf S}_{\begin{ytableau} \ \\ \ \end{ytableau}} = \frac{1}{(\H_1 - \H_2)(\H_1 - \H_3)} \Big( \H_1^2 \, {\bf P}_{1,1}^2 + \H_1^3 \, {\bf P}_{2,1} - \sigma_3 \, {\bf P}_{2,2} \Big)
    }
\end{equation}
These formulas are direct analogue of \eqref{u v 2 level B = 1} and \eqref{3-Schur 2 level B=1} for different choice of normalization of $A_1 B_1$. Despite the homogeneous formulas look simpler, the rules for reduction to the Jack case remains unclear. This approach could be continued to higher level, but we do not provide it here. 

The main advantage of homogeneous approach is that one can restore the dependence on parameters $\H_i$ of the unknown functions $A,B,u,v,w$ of the cut-and-join operator using degree arguments. 

\section{Conclusion}

In this paper we provide a very clear and simple prescription to generate the $3$-Schur functions.
Namely, we explicitly define the cut-and-join operator $\hat W_2=\hat \psi_3$ through time variables
in \eqref{3SchursW2}.
Together with the Pieri-rule generators $\hat e_0 = { \bf P}_{1,1}$ and $\hat f_0 = \frac{\p}{\p {\bf P}_{1,1}}$
this is enough to recursively generate all other $\hat e_{k+1}$
and the $3$-Schur functions, which are the common eigenfunctions of all operators $\hat \psi_k$
of the Yangian.
In practice these $3$-Schur functions are associated with all plane partitions tableaux
and for a given plane partition they differ by the easily controlled time-{\it in}dependent factors.
 
What remains to be fixed are the values of coefficients $A,B,u,v,w$ in \eqref{3SchursW2},
which are just trivial in the case of $2$-Schur and Jack functions.
We demonstrated that for $3$-Schurs they are {\it non}-trivial,
but more calculations or alternative insights are needed to fix them
and understand whether one needs additional terms in \eqref{3SchursW2} or not.
Just like in the previous attempts in \cite{Morozov:2018fjb,Morozov:2018fga, Morozov:2018lsn}
the problems seem to mount up at the first essentially 3-dimensional {\it level four},
where the formulas of \cite{Wang3} also do not look simple and transparent.
Deeper analysis and the relation to Yangian representation theory and free-field representation program \cite{DiffOper1Morozov:2021hwr, DiffOper2Morozov:2022ocp} will be presented elsewhere.

The four obvious possible reasons for the partial failure of our approach at level 4 are
\begin{itemize}
    \item computational mistake
    \item omission of some important contributions to the cut-and-join operator $\hat W_2$ (like changing $i$ by two)
    \item an overoptimistic/naive assumption that $\hat W_2$ looks natural in variables ${\bf P}_{a,i}$
    \item erroneous choice of the constants $C_{\bar \pi}$.
\end{itemize}
 
We do not dwell upon the first three options, but provided an explicit list of $C_{\bar \pi}$ in Section \ref{C consts},
which we used  – so that an interested reader can check the last option. \\

We {\it hope} that our brief but detailed exposition of the problem will help to attract
more researchers to the subject of $3$-Schur polynomials,
which remains one of the core issues for the theory of DIM algebras and
numerous string models with Yangian and DIM symmetries.
 
\section*{Acknowledgements}

We are indebted  for illuminating discussions to D.Galakhov, A.Mironov and V.Mishnyakov. We are grateful to the anonymous reviewer for suggestion of elegant formula \eqref{ratio of C}.

Our work is supported by the Russian Science Foundation (Grant No.20-71-10073).

\printbibliography

\end{document}